# Magnetic properties and phase diagrams of the ferrimagnetic triangular nanotube with core-shell structure: A Monte Carlo study

Maen Gharaibeh[1,*], Samah Alqaiem[1], Abdalla Obeidat[1], Ahmad Al-Qawasmeh[2], Sufian Abedrabbo[2] and Mohammad H. A. Badarneh[3]

[1]Department of Physics, Jordan University of Science and Technology, Irbid 22110, Jordan
[2]Department of Physics, |Khalifa University, Abu Dhabi, UAE
[3]Science Institute, University of Iceland, 107 Reykjavík, Iceland

*Corresponding author: magh@just.edu.jo

**Abstract**

Monte Carlo simulation has been employed to investigate the magnetic properties and phase diagrams of ferrimagnetic mixed-spin (½, 1) triangular Ising nanotube with core-shell structure. In particular, the effect of the exchange couplings and the temperature on the magnetic and thermodynamic properties, hysteresis loops as well as the compensation temperature have been discussed in detail. Moreover, the effects of the single-ion anisropy, as well as external magnetic field, have been examined. The threshold values of the exchange couplings and single-ion anisotropy have been found, determining whether the system exhibits a compensation temperature. We have found that the appearance of the compensation temperature is strongly linked with the system parameters. Moreover, we have obtained the double and triple hysteresis loops for certain physical parameters in the considered magnetic system.

**Keywords:** 2D-Lattice, Monte Carlo Simulation, Ising Model, Magnetic Properties, Phase Diagrams, Hysteresis Loops, Crystal Field.

## 1. Introduction

Magnetic nanotubes are a promising candidate for various innovative technological applications [1-4], owing to their enhanced magnetic anisotropy due to their extended length and nanoscale lateral size [5]. Moreover, nano-sized magnetic particles, e.g., nanoparticles, nanowires, and nanotubes [6-11], have received much attention due to their immense applications in technology [12-14]. In addition, these nano-size magnetic particles can be used in spintronics devices [15], drug delivery [16], sensors [17], and bio-molecular motors [18].

Experimentally nanoparticles are fabricated with techniques such as the atomic layer deposition [19-21]. From a theoretical point of view, the magnetic and thermodynamics properties of those nanoparticles have been studied using different methods such as the effective field theory (EFT) [22-24] and Monte Carlo simulations [25-28]. On the other hand, the ferrimagnetic mixed spin Ising system has attracted a lot of interest and become the subject of various statistical mechanics and solid-state physics studies. Generally, these models are implemented to study the critical behavior [29], the magnetic properties [30], and the compensation temperatures [31, 32] of various spins structures [33-36]. Among these models, the Ising model used in this work is the simplest, yet it has proven to be very effective in predicting the magnetic properties of ferromagnetic materials. In addition, the Ising model has proved to be able to predict some unusual behavior of several mixed spin structures [37, 38].

Among the different mixed spin structures studied within the Ising model, the two-dimensional structure remains the most widely studied structure in the literature. Various shapes of the two-dimensional structures such as the triangular [39], cubic [40], honeycomb [41] have been studied thoroughly within this model. In recent years, a growing interest has been observed in simulating the magnetic properties of various shapes of the two-dimensional nanotube structure [42-44] within the framework of the Ising model because of its wide range of applications in high-density recording media.

Effective field theory with correlation has been used to investigate the magnetic properties of cylindrical Ising nanowire or nanotube with diluted surface [45, 46]. Also, it has been used to study the phase diagrams of a monodomain ferroelectric nanoparticle described by the transverse Ising model, where the nanoparticle consists of concentric hexagonal rings [47]. In addition, EFT has been used to investigate the hysteresis loops of a ferromagnetic and ferrimagnetic three-walled mixed-spin Ising nanotube with an inner hexagonal vacancy [48].

Theoretical simulations based on Monte Carlo method were used to investigate the magnetic and thermodynamic properties of nanotubes. Among them, spin-1/2 and spin-1 Ising model on a nanotube structure [49], mixed spin-1/2 Ising model and spin-1 Blume-Capel model in a hexagonal nanotube two layers system [50], a cylindrical a ferrimagnetic Ising nanotube of the spin-1 core and shell structure [51] and a ferrimagnetic mixed spin-1/2 and spin-3/2 Ising nanoisland with a double-layer quadrangle core-shell structure [52].

Experimentally interesting magnetic behaviors have been observed in core-shell nanotubes [53-56], which motivate us to carry theoretical calculations based on Monte Carlo simulations based on the Metropolis update protocol to better understand these magnetic properties core-shell nanotubes. Thus, in this work, we use Monte Carlo simulation within the framework of the Ising model to study several magnetic properties of the core-shell triangular nanotube structure, considering the effect of anisotropy interaction and the interaction of different spins with an external magnetic field. More specifically, we aim to investigate the effect of the exchange interaction coupling and the crystal field on the thermal magnetization, critical and compensation behavior, susceptibility, specific heat, hysteresis loop, and phase transition diagrams on a two-dimensional nanotube structure composed of spin -1 and spin -1/2 within the framework of Monte Carlo simulations. The existence of the compensation temperature in the considered triangular nanotube makes it a potential candidate for use in the area of thermomagnetic data storage and magneto-optical recording media devices [57-59].

## 2. Model and formalism
### 2.1 Lattice structure and Hamiltonian

The nanotube structure studied in this work is shown in Fig. 1. The structure consists of $L$ layers, each with two different sublattices, the sublattice S, which consists of $S$-type of spins -1 is on the tube's shell, and the sublattice C with $\sigma$ -type of spins -1/2 is in the core of the tube.

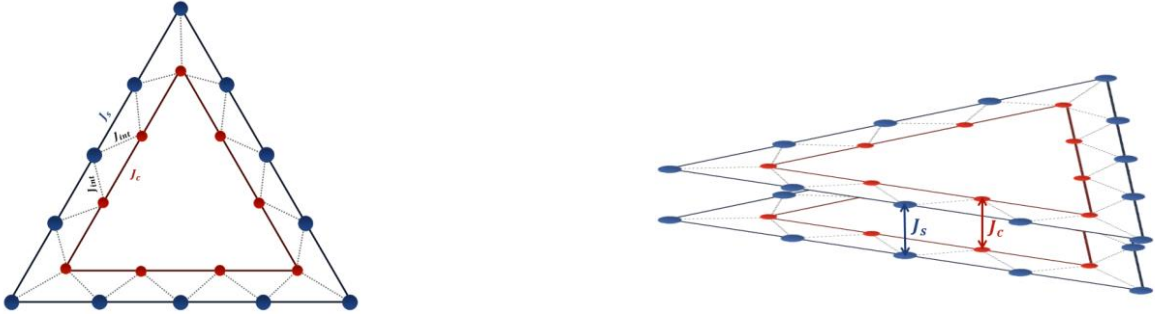

**Fig.1**: Systematic representation of two different prospects of the nanotube structure. The left-hand side figure shows the nanotube cross-section, while the right-hand side figure shows the different layers of the nanotube. The red atoms at the core of the nanotube are called core atoms that have spin values ( $\sigma=\pm0.5$), while the blue atoms at the tube's shell are called shell atoms, and those have spin values ($S = 0, \pm1$).

The Hamiltonian terms consist of $S - S, \sigma - \sigma$ ferromagnetic interaction, and $S - \sigma$ ferromagnetic interaction, and spin -external field interaction and interaction of spin type S with crystal field. The Hamiltonian of the system can be written as follows:

$$H = -J_s \sum_{\langle i,j \rangle} S_i S_j - J_c \sum_{\langle l,k \rangle} \sigma_l \sigma_k - J_{int} \sum_{\langle i,l \rangle} S_i \sigma_l - B \sum_i (S_i + \sigma_i) - D_s \sum_i S_i^2 \qquad (1)$$

Where $J_s$, $J_c$, $J_{int}$ stand for the coupling constants between the spins $S - S, \sigma - \sigma$ and $S - \sigma$, respectively. The summation indices $\langle i,j \rangle$ and $\langle l,k \rangle$ denote the summation over all the nearest neighbors spins $S - S, \sigma - \sigma$, respectively. The fourth term is the interaction of the external magnetic field $B$ with spin type $S$ and spin type $\sigma$, and the last term is the interaction of the crystal field $D_s$ with spin type $S$. In our simulations, we choose a positive value for both $J_s$ and $J_c$ to ensure a ferromagnetic interaction between spins of the same type. In contrast, a negative value of $J_{int}$ is chosen to ensure ferromagnetic interaction between different spin types.

*2.2 Monte Carlo Simulation and Calculations*

The ferrimagnetic nanotube shown in fig. 1 was simulated using the Monte Carlo simulation method based on the metropolis algorithm [60]. The periodic boundary conditions were applied along the three axes x, y, and z. In this method, new configurations were generated by choosing a random spin of the S type or $\sigma$ type, and then we flip it randomly to the other possible values of the spin. Each flip will then be accepted or rejected according to the metropolis algorithm. A 1000000 Monte Carlo steps were used to equilibrate the system, followed by 850000 steps for each spin configuration. The results are reported for the system shown in fig. 1 with a total number of atoms $N = 10500$ corresponding to $L = 500$ layers and $N_s = 12$ for the shell atoms with spin type $S$ and $N_c = 9$ for the core atoms with spin type $\sigma$. Here, we fix number of atoms in the core and shell, and we vary the number of layers. Since the finite size effects might arise due to the lack of layers, we performed additional simulations with *L>500*, but no significant differences were found from the results presented here. For the error calculation, we use the method of blocks in which the *L*-size is divided into $n_b$ blocks of length $L_b = L/n_b$. The number of blocks is chosen such that $L_b$ is higher than the correlation length. Therefore, error bars are calculated by grouping all the blocks and then taking the standard deviation [61]. The magnetic properties were calculated as follows:

The magnetization per site for the shell sublattice $M_s$ and the core sublattice $M_c$ is given by:

$$M_s = \frac{1}{N_s} \langle \sum_{i=1}^{N_s} S_i \rangle, \qquad (2)$$

$$M_c = \frac{1}{N_c} \langle \sum_{l=1}^{N_c} \sigma_l \rangle. \qquad (3)$$

The average magnetization per site for the whole system is given by:

$$M = \frac{N_s M_s + N_c M_c}{N_c + N_s}. \qquad (4)$$

The magnetic susceptibilities for the shell sublattice $\chi_s$ and for the core sublattice $\chi_c$ are given by:

$$\chi_s = N_s \beta (\langle M_s^2 \rangle - \langle M_s \rangle^2), \qquad (5)$$

$$\chi_c = N_c \beta (\langle M_c^2 \rangle - \langle M_c \rangle^2). \qquad (6)$$

And the total susceptibility is

$$\chi = \frac{N_s\chi_s + N_c\chi_c}{N_s + N_c}. \tag{7}$$

Where $\beta = 1/k_BT$, $T$ is the absolute temperature, and $k_B$ is the Boltzmann factor. For simplicity, we set $k_B = 1$.

Finally, we have calculated the specific heat, $C$, of the system as follows:

$$\frac{C}{k_B} = \frac{\beta^2}{N}(\langle H^2 \rangle - \langle H \rangle^2). \tag{8}$$

The compensation temperature, $T_{comp}$, is defined as the temperature at which the magnetization of two sublattices adds up to zero given a zero-total magnetization of the lattice as given in Eq. 4. To compute $T_{comp}$ from the magnetization curves, the crossing point of the magnetizations of the two sublattices has to be determined under the following condition:

$$|M_s(T_{comp})| = |M_c(T_{comp})| \tag{9}$$

$$sign\left(M_s(T_{comp})\right) = -sign\left(M_c(T_{comp})\right) \tag{10}$$

with $T_{comp} < T_C$, where $T_C$ is the critical temperature. In this paper, $T_C$ is determined from the divergence of the susceptibilities' curves.

**3. Results and discussions**

This section will present our results for the magnetic and thermodynamical properties of the mixed-spin nanotube structure. Furthermore, the influence of the Hamiltonian parameters, $(J_s, J_c, J_{int}, D_s)$ as given in Eq. 1, on these properties will be explored thoroughly.

3.1 Magnetic properties and phase diagrams

In this section, the general trend behavior of the nanotube structure such as the total magnetization given in Eq. 4, the magnetization of the shell and core sublattices given in Eq. 2 and Eq. 3 respectively, the susceptibility and the specific heat in Eq. 7 and Eq. 8 respectively will be presented at various Hamiltonian parameters. In addition, to explore the effect of these Hamiltonian parameters on the compensation temperature, $T_{comp}$, defined in Eq. 9 and Eq. 10 and the critical temperature, $T_C$, defined as the second zero of the total magnetization, phase diagrams, and contour plots of $T_{comp}$ and $T_C$ at various Hamiltonian parameters will be presented.

The effect of $J_s$ on the sublattices magnetization, total magnetization, the susceptibility, and the specific heat is illustrated in Fig. 2 at which we fix the parameters $J_c = 2$, $J_{int} = -0.1$, $D_s = 0$ and $B = 0$ at different values of $J_s$ of $0.1, 0.3, 0.5, 0.7, 0.9$. As shown in Fig. 2(b), the system total magnetization has two zero points for $J_s \leq 0.3$ corresponding to the compensation and critical temperature, respectively; after this threshold value $J_s > 0.3$, the system exhibits only a critical temperature corresponding to the single zero of the magnetization curve. It is noticeable that below the threshold value $J_s \leq 0.3$, the compensation temperature increases with increasing $J_s$ while the critical temperature is fixed; after the threshold value $J_s > 0.3$, the compensation temperature disappears, and the critical temperature increases with increasing $J_s$.

As shown in Fig. 2(a), the shell and the core sublattices start from a fully ordered state at low temperature corresponding to magnetization value of -1 and 0.5 for the shell and core sublattices, respectively, and the magnetization for each sublattice starts to change as the temperature increases due to the increase in the disorder in each sublattice. For $J_s \leq 0.3$, the magnetization for the shell and core sublattices satisfies Eq. 8 and Eq. 9 at the compensation temperature where sublattices magnetization has opposite signs and equal magnitude while for $J_s > 0.3$ Eq. 8 and Eq. 9 are not satisfied for all temperature values, which shows that the compensation temperature is absent at these values of $J_s$.

The total specific heat and susceptibility of the system shown in Fig. 2(c) and Fig. 2(d) have two peaks for $J_s \leq 0.3$ corresponding to the compensation and critical temperature, respectively. The position of the first peak is fixed, and the position of the second peak increases with increasing $J_s$. After the threshold value $J_s > 0.3$, both the total specific heat and susceptibility curves exhibit one peak at the critical temperature, and the value of the peak increases with increasing $J_s$. Moreover, one can notice that all the internal energy curves have the same behavior, i.e., the system's internal energy increases with increasing the temperature. For $J_s \leq 0.3$ the internal energy curves approach two discontinuity points, where the first (second) discontinuity point coincides with the location of the compensation temperature (critical temperature). On the other hand, for $J_s > 0.3$ the internal energy curves approach only one discontinuity point at which a second-order phase transition to paramagnetic phase occurs in the system. It is worth mentioning that the existence of the first discontinuity point for $J_s \leq 0.3$ is due to an abrupt drop in the magnetization of the shell sublattice in the low temperatures, as shown in Fig. 2(a). The inset in Fig. 2(d) shows the location of the first peak in the susceptibility curves for $J_s \leq 0.3$. We can remark that the same behavior for the magnetization, specific heat, internal energy and susceptibilities curves have been reported in Refs. [51, 62-65].

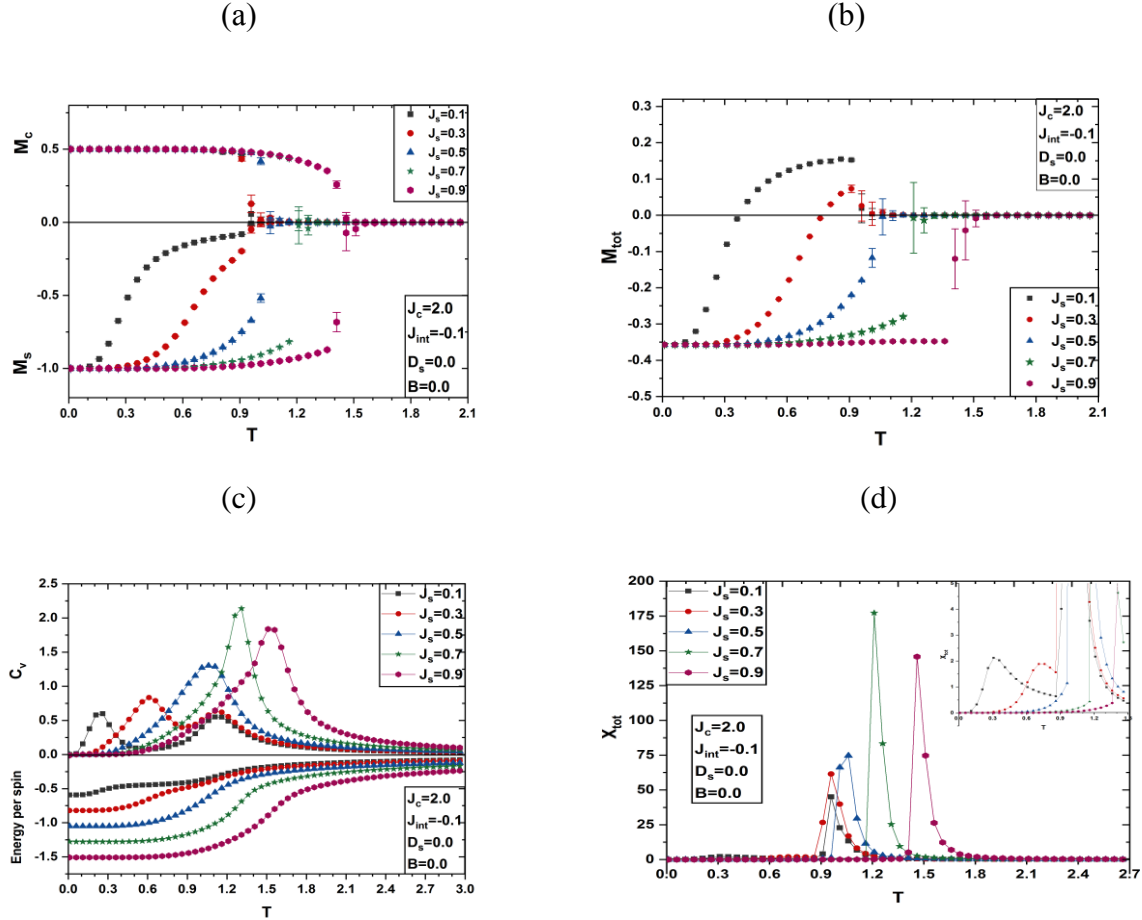

**Fig. 2.** The temperature dependencies of (a) sublattice magnetization, (b) total magnetization, (c) specific heat and the energy per spin where the peaks in the specific heat curves correspond to the inflection points in the energy per spin curves, (d) total susceptibilities, for $J_c = 2, J_{int} = -0.1, B = 0, D_s = 0$ for different values of $J_s = (0.1, 0.3, 0.5, 0.7, 0.9)$.

Similarly, we investigate the effect of $J_c$ on the sublattices magnetization, total magnetization, the susceptibility, and the specific heat as shown in Fig. 3 at which $J_s$ is fixed at a value of 0.1 where the system exhibits both compensation and critical temperatures, $J_{int}$ is fixed at -0.1 and both $D_s$ and $B$ are fixed at zero, and the $J_c$ is varied from 2 to 4 by 0.5 steps. As shown in Fig. 3(b), for all $J_c$ values, the total magnetization curves of the lattice have two zero points corresponds to the compensation and critical temperatures, respectively. The position of the first zero point is fixed while the position of the second zero increases with increasing $J_c$, which means that the system compensation temperature remains constant for all $J_c$ values while the critical temperature increases with $J_c$.

As shown in Fig. 3(a), the magnetization of the core and shell sublattices starts from the values 0.5 and -1 respectively at low temperature where both sublattices are in full ordered state, and they start to change as the disorder in each sublattice increase with increasing the temperature. For all values of $J_c$, both $M_s$ and $M_c$ satisfy Eq. 8 and Eq. 9, which ensures that the system manifests compensation behavior for all the values of $J_c$.

As shown in Fig. 3(c) and Fig. 3(d), respectively, the specific heat and susceptibility exhibit two peaks at the compensation temperature and the critical temperature. The first peak position remains fixed for values of $J_c$, which confirms that the system compensation temperature remains constant, while the position of the second peak increases with $J_c$, which confirms that the system critical temperature increases with $J_c$. In addition, the internal energy of the system increases with increasing temperature, as shown in Fig. 3(c). All the internal energy curves approach two discontinuity points, where the first (second) discontinuity point coincides with the location of the compensation temperature (critical temperature). It is worth mentioning that the existence of the first discontinuity point is due to an abrupt drop in the magnetization of the shell sublattice in the low temperatures, as shown in Fig. 3(a). The inset in Fig. 3(d) shows the location of the first peak in the susceptibility curves.

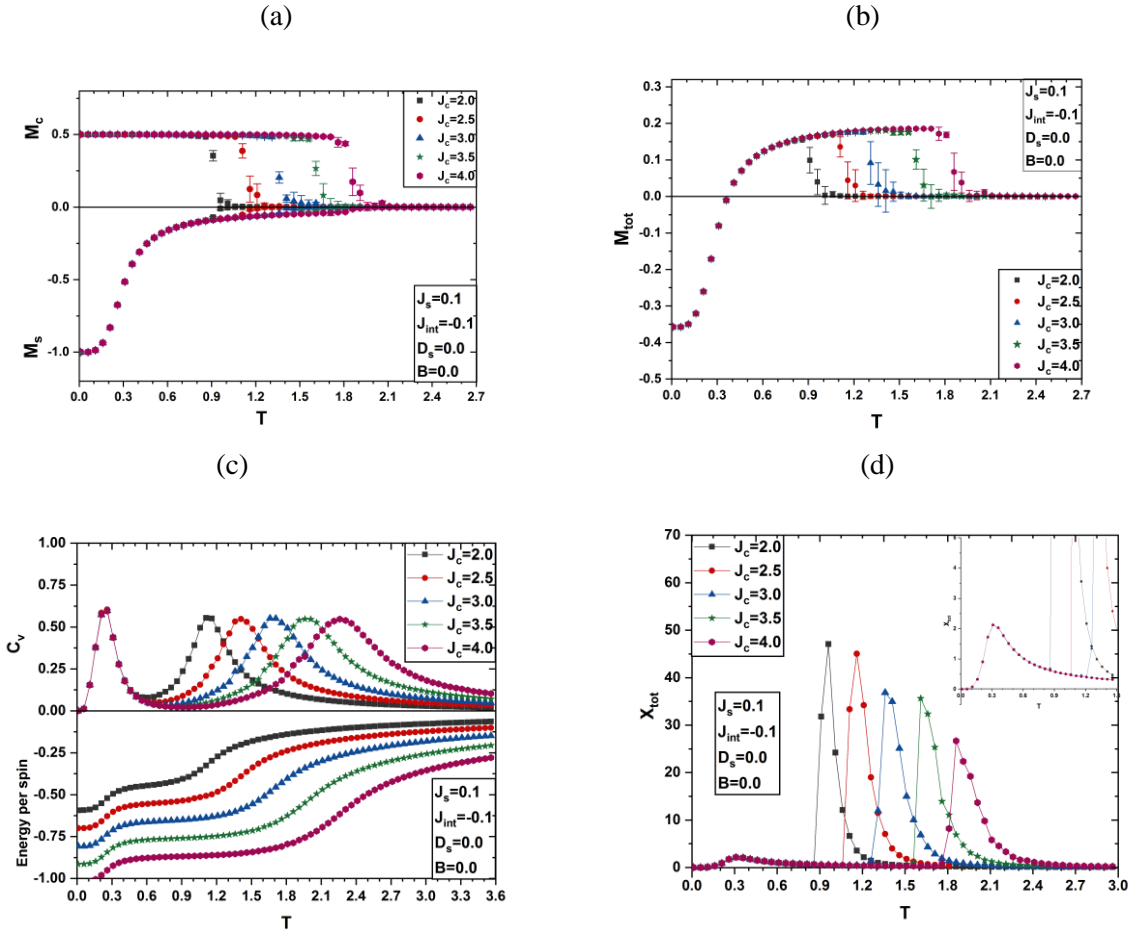

**Fig. 3.** The temperature dependencies of (a) sublattice magnetization, (b) total magnetization,

(c) specific heat and the energy per spin where the peaks in the specific heat curves correspond to the inflection points in the energy per spin curves, (d) total susceptibilities, for $J_s = 0.1, J_{int} = -0.1, B = 0, D_s = 0$ for different values of $J_c = (2.0, 2.5, 3.0, 3.5, 4.0)$.

To better illustrate the effect of $J_s$ and $J_c$ on the compensation temperature of the system, it is useful to plot the calculated value of $T_{comp}$ at various values of $J_s$ and $J_c$ and fixed $J_{int} = -0.1, D_s = 0$ as shown in Fig. 4. Here we vary $J_s$ from 0.1 to 1 in 0.1 steps, and at each value, we vary $J_c$ from 0.4 to 4 at 0.4 steps. It is clear from the figure that the system exhibits a compensation behavior for $J_s \leq 0.9$ and $J_c \geq 1.4$. It is also quite noticeable that as we increase the value of $J_s$, the range of $J_c$ values where the compensation temperature exists decreases.

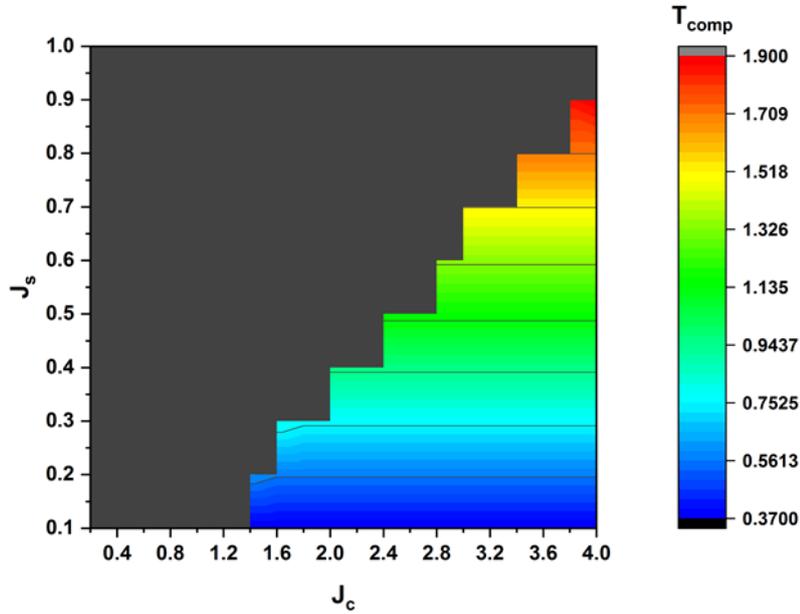

**Fig. 4.** A Contour plot of the compensation temperature of the nanotube system at different values of $J_s$ and $J_c$ with fixed parameters $D_s = 0, J_{int} = -0.1$.

The effect of $J_{int}$ on the sublattices magnetization, total magnetization, the susceptibility, and the specific heat has been investigated as well as shown in Fig. 5 at which the parameters $J_s, J_c, D_s$, and $B$ are fixed at the values of 0.1, 2.5, 0, 0 respectively where both the compensation and critical temperature of the system exist as discussed in Fig. 4 and the $J_{int}$ values are varied from -2.0 to -0.1. The effect of $J_{int}$ on the system compensation and critical temperatures is quite apparent. As shown in Fig. 5(b), for $J_{int} \geq -0.3$ the magnetization curves have two zero points at the compensation and critical temperature, respectively, and below this threshold value where $J_{int} < -0.3$ the magnetization curves exhibit a single zero point at the critical temperature. This indicates that the system exhibits both compensation and critical behavior only for $J_{int} \geq -0.3$. It is also noticeable that both the critical and compensation temperature for the system increases with decreasing $J_{int}$ as the first and second zeros of the magnetization curves shift to the right with decreasing $J_{int}$.

This result can be confirmed further by analyzing the total specific heat and susceptibility curves shown in Fig. 5(c) and Fig. 5(d), respectively. Here, two peaks are observed for both curves for $J_{int} \geq -0.3$, which occurs at the compensation and critical temperature, respectively, and one peak for $J_{int} < -0.3$, which occurs at the critical temperature, which ensures that the system exhibits a compensation behavior only for $J_{int} \geq -0.3$. The position of these peaks moves to the right as we decrease $J_{int}$, which confirms that both the compensation and critical temperature increases with decreasing $J_{int}$. The inset of Fig. 5(d) shows the location of the first peak in the susceptibility curves for $J_{int}$ less than or equal to -0.3. Similar behavior has also been observed in Refs. [63, 64, 66, 67].

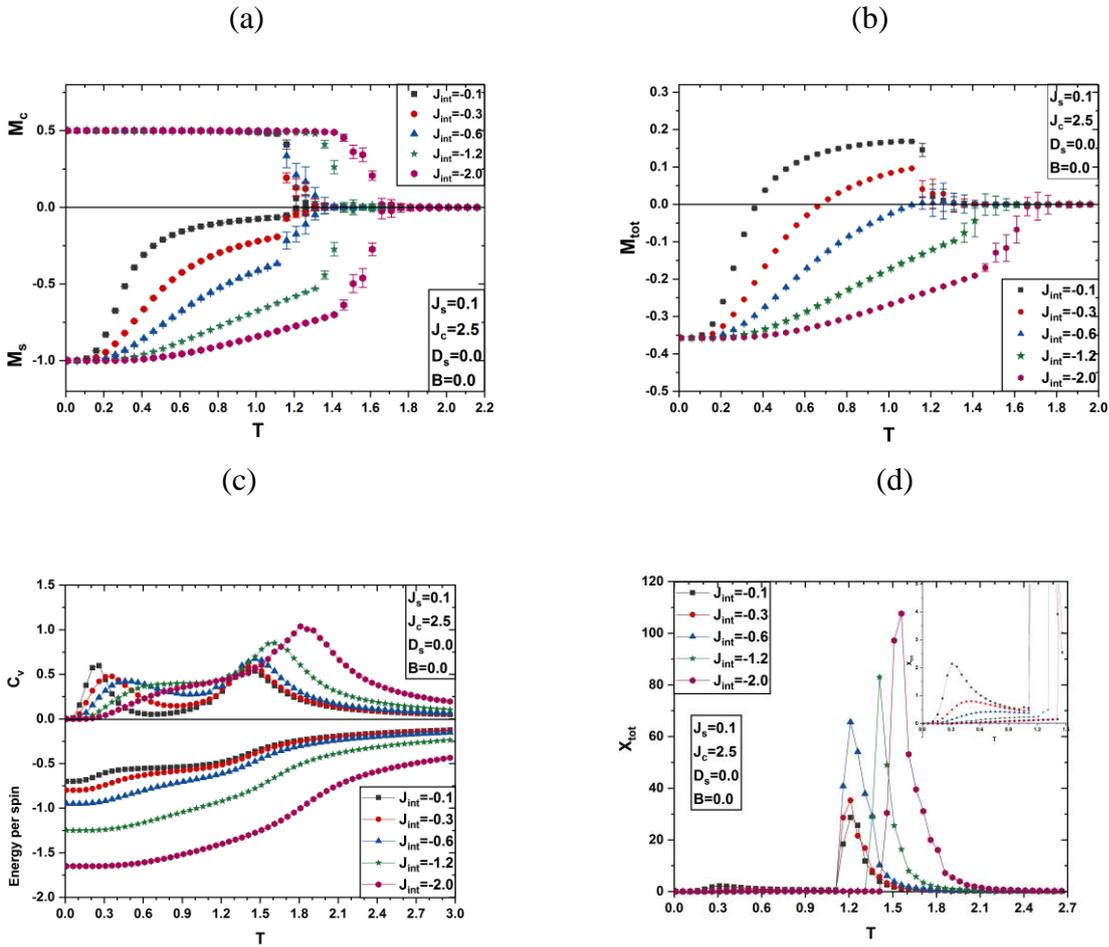

**Fig. 5.** The temperature dependencies of (a) sublattice magnetization, (b) total magnetization, (c) specific heat and the energy per spin where the peaks in the specific heat curves correspond to the inflection points in the energy per spin curves, (d) total susceptibilities, for $J_s = 0.1, J_c = 2, B = 0, D_s = 0$ for different values of $J_{int} = (-0.1, -0.3, -0.6, -1.2, -2.0)$.

The influence of the crystal field $D$ on the total magnetization of the system is shown in Fig. 6. here, we fix the parameters $J_c = 2.0, J_{int} = -0.1, B = 0$ for selected values of $J_s$=0.1,0.3,0.5,0.7,0.9 and we vary $D_s$ from -2.0 to 0 in 0.5 steps.

As shown in Fig. 6(a), for $J_s = 0.1$, the magnetization curves have two zero points at the compensation and critical temperatures, respectively, only in the absence of the crystal field $D_s = 0$, for all other values of the crystal field, the magnetization curves have one zero point at the critical temperature. This means that the system possesses a compensation behavior only in the absence of the crystal field. It is worth noting that the critical temperature appears to be insensitive to the crystal field in this case since the zero of the magnetic field curves remains almost unchanged for all $D_s$ values.

For $J_s = 0.3$ shown in Fig. 6(b), the system possesses a compensation behavior only for $D_s = 0, -0.5$ where the magnetization curve has two zero points. The position of the first zero point, which corresponds to the compensation temperature, increases significantly with decreasing $D_s$. For all other $D_s$ values, the system exhibits a critical behavior only corresponding to the single zero point of the magnetization curves. The position of the critical temperature seems to be fixed with changing $D_s$ since all the corresponding magnetization curves cross the axis at the same point.

For $J_s = 0.5$ shown in Fig. 6(c), the system exhibits a compensation behavior only for $D_s = -1$ corresponding to the first zero point of the magnetization curve. For all other $D_s$ values, the system exhibits only a critical temperature where the magnetization curves have only one zero point. The critical temperature value seems to be more impacted by the change in $D_s$ than for the cases of $J_s = 0.1, 0.3$ where we observe a more significant shift to the right of the magnetization curve zero point as the value of $D_s$ decrease.

For $J_s = 0.7$ and $J_s = 0.9$ shown in Fig. 6(d) and Fig. 6(e), respectively, the system does not have a compensation temperature for all $D_s$ values as all the corresponding magnetization curves have a single zero point. It is quite noticeable that the critical temperature, which corresponds to the zero point of the magnetization curve, increases more significantly as we decrease $D_s$ than for the cases of $J_s = 0.1, 0.3, 0.5$. Similar phenomena are observed in Refs. [51, 64, 67, 68]

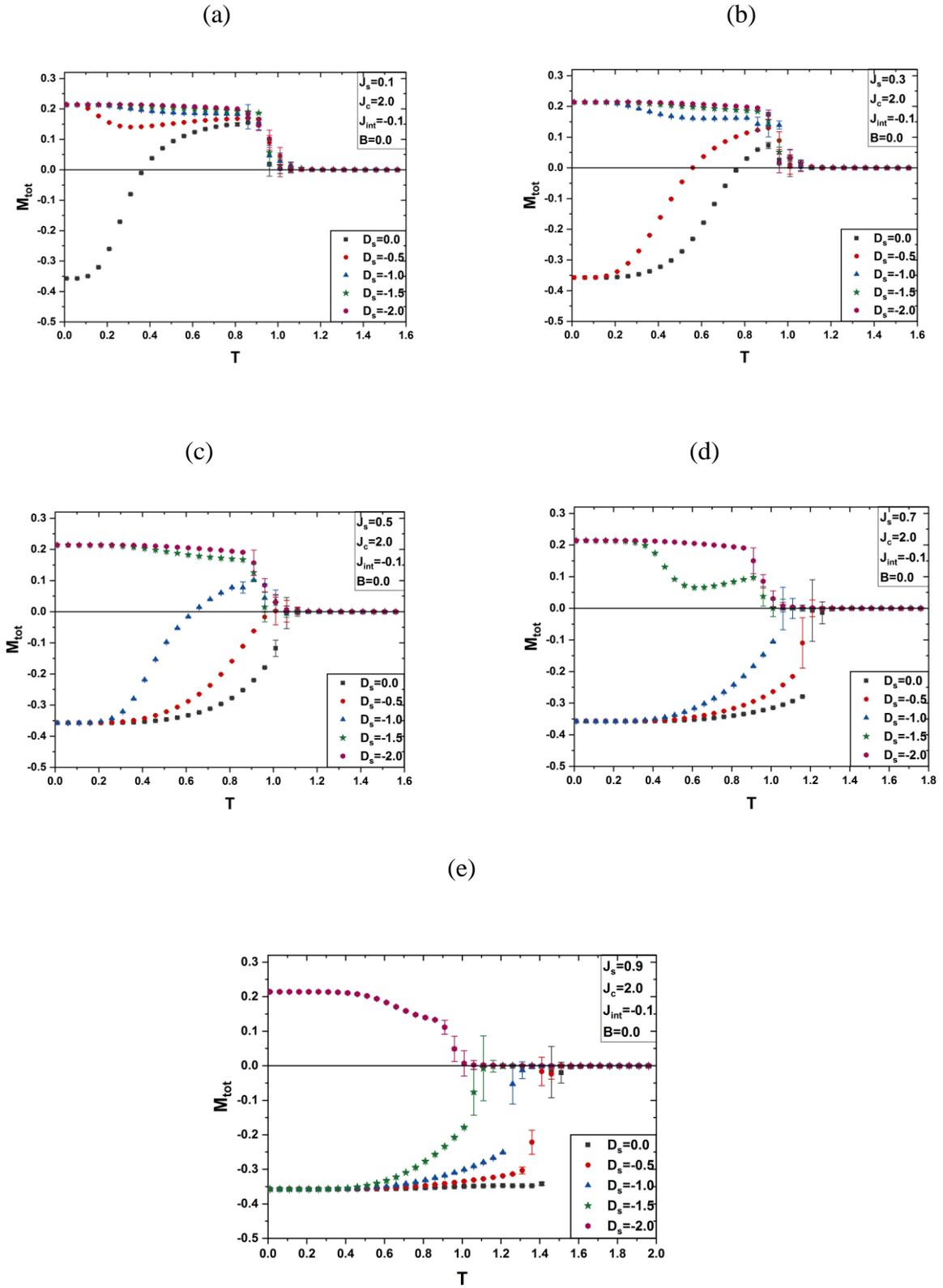

**Fig. 6.** The temperature dependencies of total magnetization at various crystal field values at fixed parameters (a) $J_s = 0.1, J_c = 2.0, J_{int} = -0.1, B = 0$ (b) $J_s = 0.3, J_c = 2.0, J_{int} = -0.1, B = 0$ (c) $J_s = 0.5, J_c = 2.0, J_{int} = -0.1, B = 0$ (d) $J_s = 0.7, J_c = 2.0, J_{int} = -0.1, B = 0$ (e) $J_s = 0.9, J_c = 2.0, J_{int} = -0.1, B = 0$.

To better illustrate the effect of $D_s$ on the compensation temperature of the system, it is useful to plot the calculated value of $T_{comp}$ at various $J_s, J_c$, and $D_s$ as shown in Fig. 7. Here, we fix $J_s$ at 0.1, 0.5, and 1 in Fig. 7(a), Fig. 7(b), and Fig. 7(c), respectively, and we vary $J_c$ and $D_s$ at $B = 0$. As shown in the figure, the range of $J_c$ values at which the system $T_{comp}$ exist decrease as $D_s$ increase. It is also worth noting that the value of $T_{comp}$ increases as the values of both $J_c$ and $D_s$ increase.

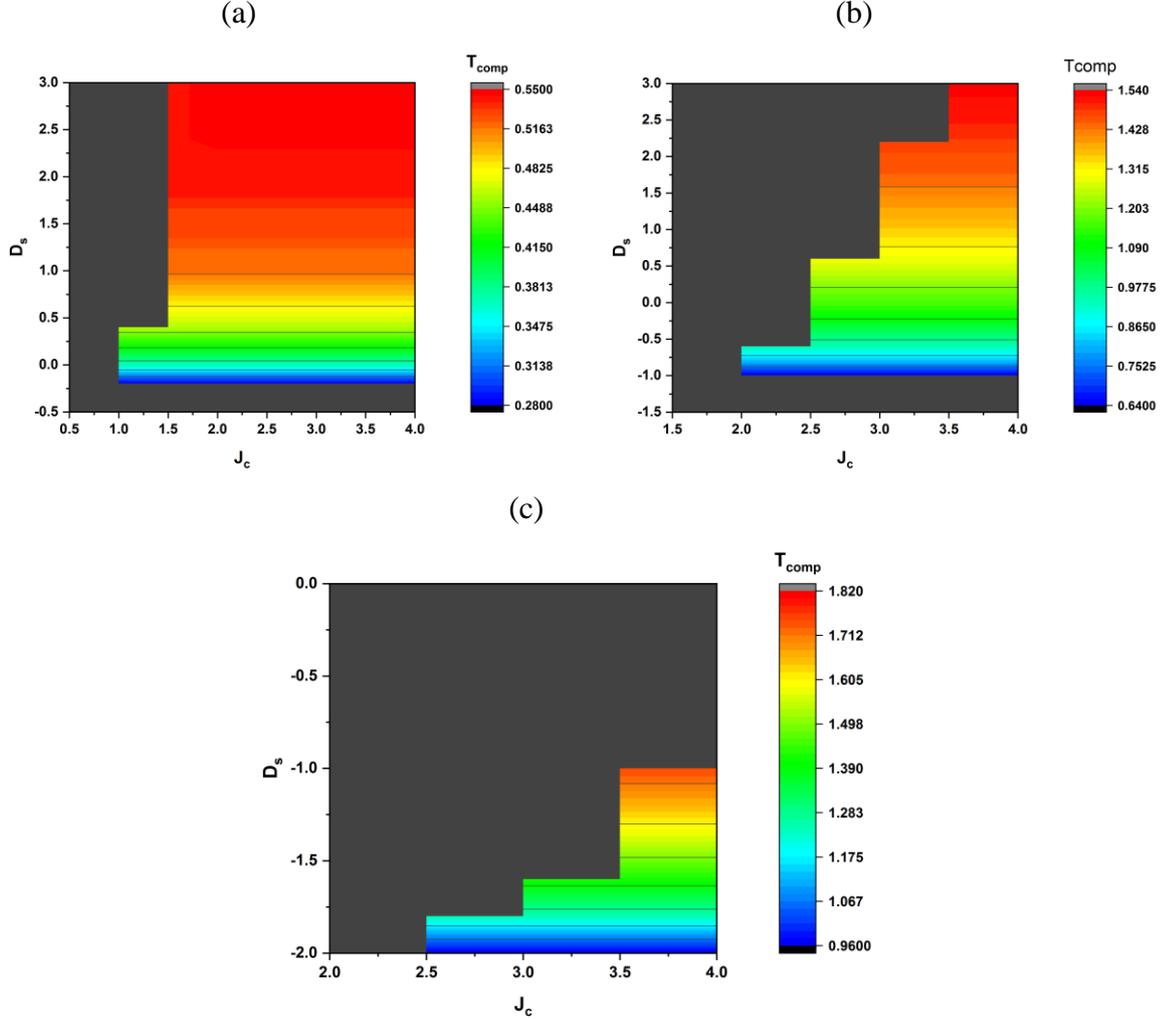

**Fig. 7.** A Contour plots of the compensation temperature of the nanotube system at different values of $J_c$ and $D_s$ and fixed $B = 0$ and fixed (a) $J_s = 0.1$ (b) $J_s = 0.5$ (c) $J_s = 1.0$.

3.2 Hysteresis loops

In this part, the effect of the Hamiltonian parameters $(J_s, J_c, J_{int}, D_s)$ on the magnetic hysteresis cycles of the nanotube structure will be investigated thoroughly.

First, we start by investigating the temperature effect on the magnetic hysteresis cycles, as shown in Fig. 8. Here we fix the parameters $J_s, J_c, J_{int}$, and $D_s$ at 0.1, 2.5, -0.1, and 0, respectively, at various temperatures of 0.3, 0.5, 0.8, and 1.1. As shown in the figure, the hysteresis cycle's surface decreases as the system temperature increases. The magnetic domains start to appear in the system as the temperature increases since the loop's area decreases until it disappears completely, and therefore less work is needed. Also, it is quite noticeable that the system transition to a superparamagnetic state occurs at a high temperature of $T = 1.1$. It should be highlighted that similar behavior has been reported in Ref. [51].

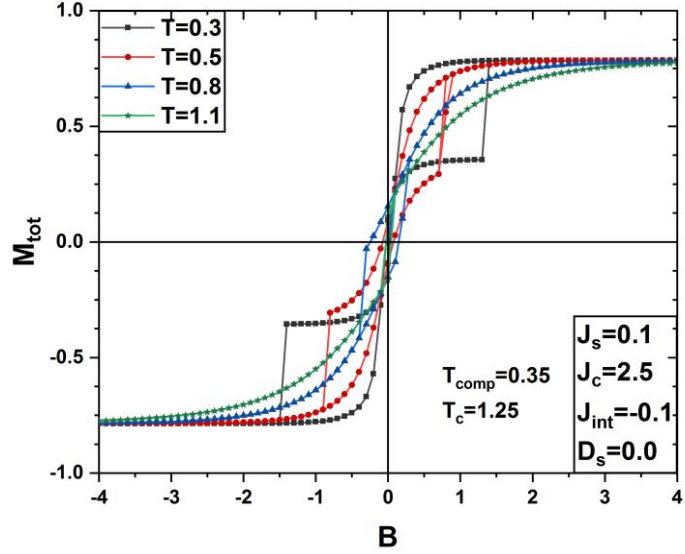

**Fig. 8.** The magnetic hysteresis cycles for the system total magnetization At different temperatures $T(T = 0.3, 0.5, 0.8, 1.1)$ and fixed parameters $J_s = 0.1, J_c = 2.5, J_{int} = -0.1$ and $D_s = 0$.

The effect of the exchange coupling parameter $J_s$ on the system hysteresis loops is shown in Fig. 9. Here, the total magnetization and sublattice magnetization with different magnetic field is presented In the left and right panels respectively at different $J_s$ values and fixed parameters $J_c = 2.5, J_{int} = -0.1, D_s = 0.0$ and temperature $T = 0.5$. From the figure, we can observe that increasing the value of $J_s$ results in a significant increase in the loop area. Also, as the value of $J_s$ increases a two steps hysteresis loops can be observed more clearly. We can remark that a similar behavior of the effect of $J_s$ on the hysteresis loops has been reported in Ref. [51, 69].

The effect of the exchange and coupling parameter $J_c$ on the total magnetization and sublattice magnetization hysteresis loops is presented in Fig. 10(a) and Fig. 10(b), respectively. Here the values of the parameters $J_s, J_{int}, D_s,$ and $T$ are fixed at 0.3, 0.1, 0, and 0.5, respectively, and the value of $J_c$ is varied from 2 to 4 in step 1. As shown the figure, the loop's area increases as the value of $J_c$ increases. However, in comparison to Fig. 9, we note that the hysteresis loop area is less sensitive to the variation of $J_c$ than the variation of $J_s$.

The effect of the exchange and coupling parameter $J_{int}$ on the total magnetization and the sublattice magnetization hysteresis loops is illustrated in Fig. 11(a) and Fig. 11(b), respectively. Here $J_{int}$ is varied at different values $(J_{int} = -0.1, -0.2, -0.4)$ at fixed $J_s = 0.3, J_c = 2.5, D_s = 0$ and temperature $T = 5.0$.

As shown in the figure, the width of the sublattices and total magnetization loops increases as we decrease $J_{int}$. Also, it is quite interesting that three-step hysteresis loops are present at low values of $J_{int}$ ($J_{int} = -0.3, -0.4$). Such behavior has been reported in Refs.[51, 62].

The crystal field $D_s$ effect on the total magnetization and sublattice magnetization is shown in the left and right panels, respectively of Fig. 12 and Fig. 13. In Fig. 12, the parameters $J_s, J_c, J_{int},$ and $T$ are fixed at 0.3, 2.5, -0.1, and 0.5, respectively, and $D_s$ takes the values -2.5, -1.5, 0. In Fig. 13, the parameters $J_s, J_c, J_{int},$ and $T$ are fixed at 0.6, 2.0, -0.1, and 0.5, while $D_s$ takes the values -2, -1, 0. As shown in both figures, as the value of $D_s$ increases, the width of the sublattices hysteresis loops increases, and consequently, the width of the total magnetization hysteresis loops increases as well. Also, it is quite noticeable that the magnetization reveals two-steps hysteresis loops at large values of $D_s$. Similar behavior has been reported in Refs. [51, 64, 69].

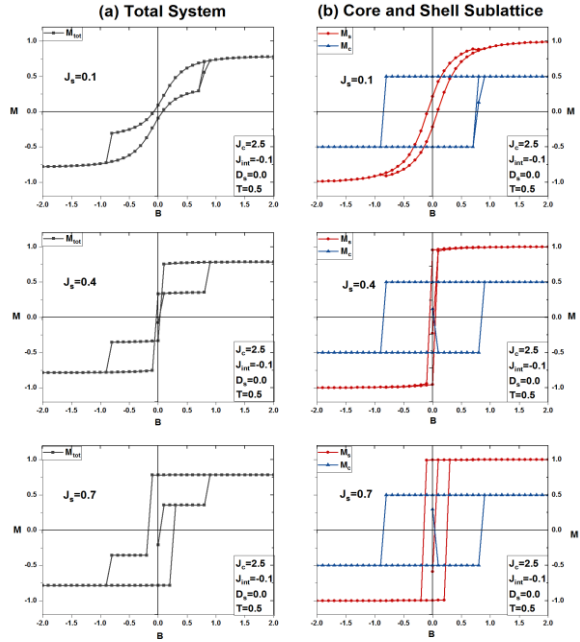

**Fig. 9.** The magnetic hysteresis cycles for different exchange coupling $J_s$ ($J_s = 0.1, 0.4, 0.7$) for (a) total magnetization (b) the core and shell sublattices Magnetizations at fixed parameters $J_c = 2.5, J_{int} = 0.1, D_s = 0, T = 0.5$.

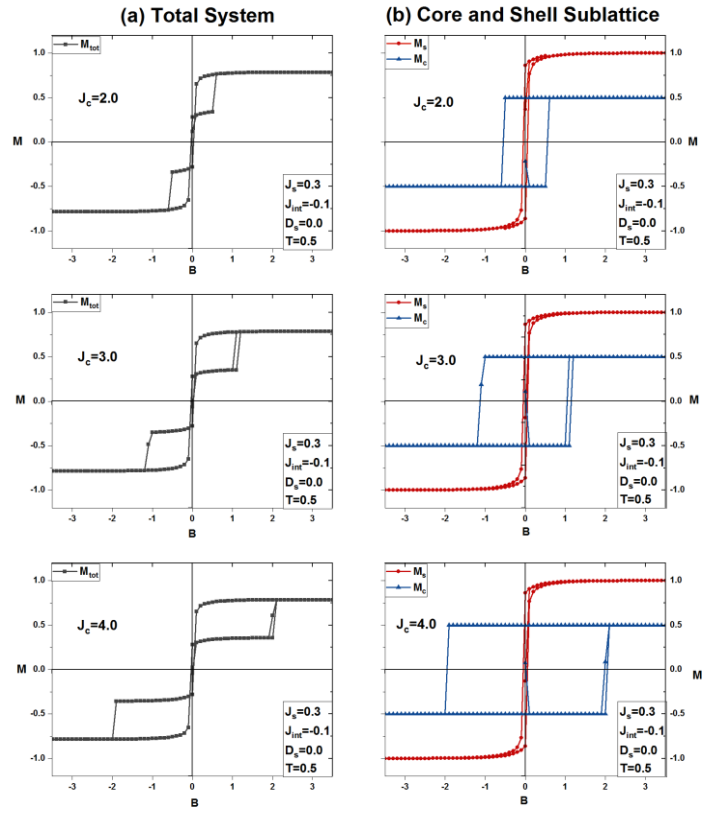

**Fig. 10.** The magnetic hysteresis cycles for different exchange coupling $J_c$ ($J_s = 2, 3, 4$) for (a) total magnetization (b) the core and shell sublattices Magnetizations at fixed parameters $J_s = 0.3, J_{int} = -0.1, D_s = 0$ and temperature $T = 0.5$.

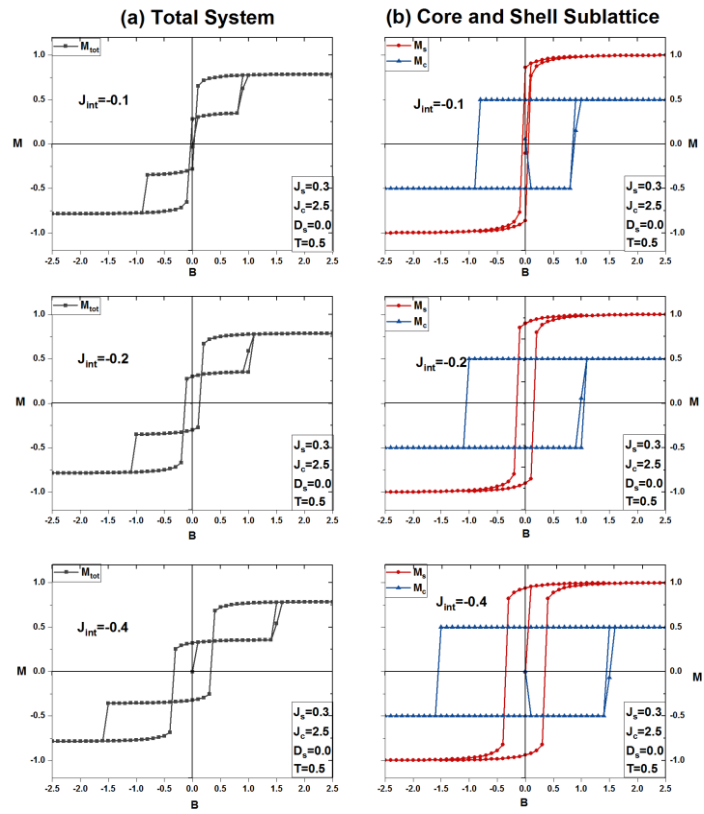

**Fig. 11.** The magnetic hysteresis cycles for different exchange coupling $J_{int}$ for (a) total magnetization (b) the core and shell sublattice Magnetizations at fixed parameters $J_s = 0.3, J_c = 2.5, D_s = 0$ and temperature $T = 0.5$.

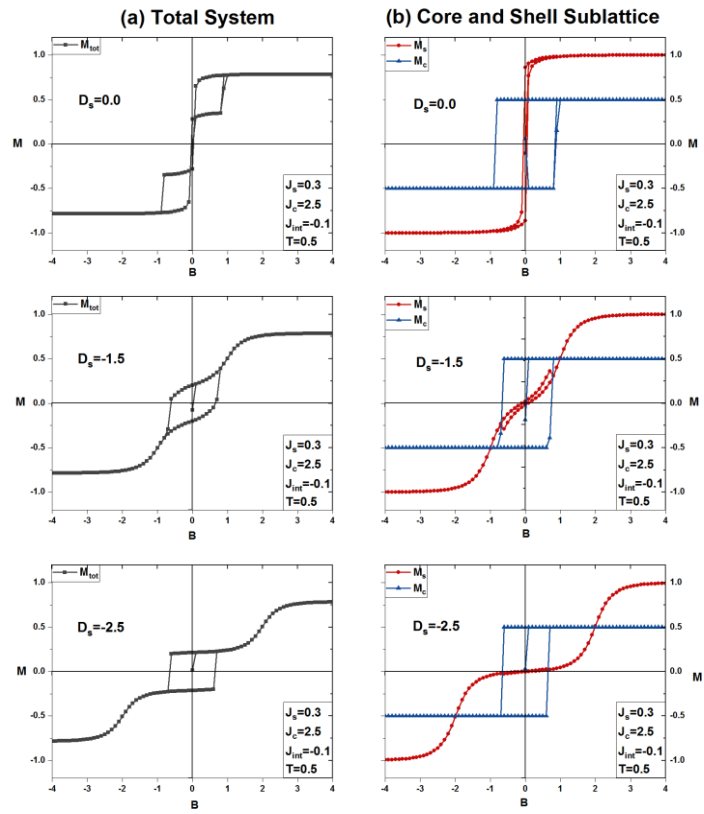

**Fig. 12.** The magnetic hysteresis cycles for different $D_s$ values $(D_s = -2.5, -1.5, 0)$ for (a) total magnetization (b) the core and shell sublattice Magnetizations at fixed parameters $J_s = 0.3, J_c = 2.5, J_{int} = -0.1$ and temperature $T = 0.5$.

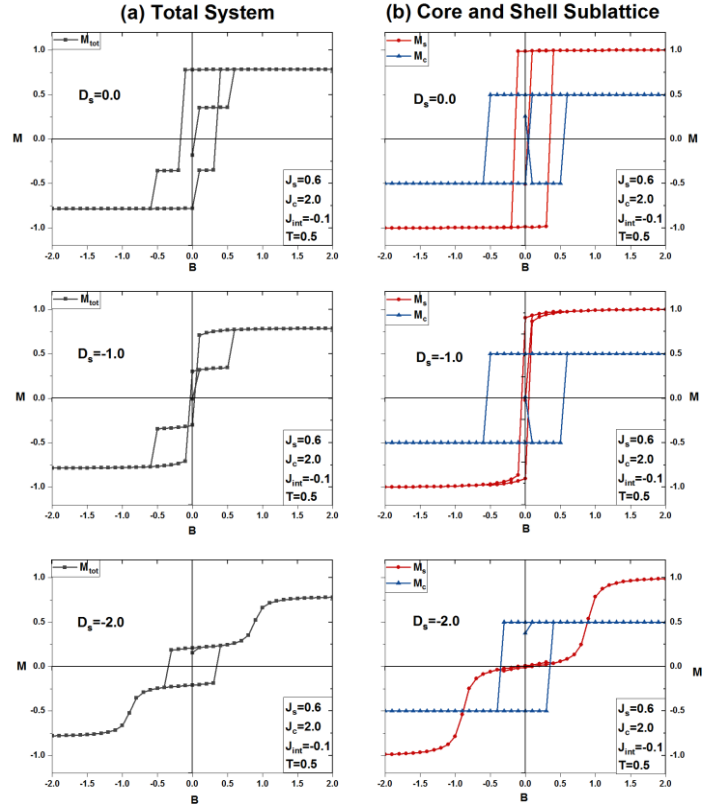

**Fig. 13.** The magnetic hysteresis cycles for different $D_s$ values $(D_s = -2.0, -1, 0)$ for (a) total magnetization (b) the core and shell sublattice Magnetizations at fixed parameters $J_s = 0.6, J_c = 2.0, J_{int} = -0.1$ and temperature $T = 0.5$.

## 4. Conclusions

In this work, Monte Carlo simulations within the framework of the Ising model have been implemented to study the effect of the Hamiltonian parameters on the magnetic properties in general and the compensation and critical behavior of the ferrimagnetic triangular nanotube. Our results suggest that the system compensation and critical temperatures are very sensitive to the exchange and coupling parameters $J_s$ and $J_c$. We observe that the system exhibits a compensation behavior only for $J_s \leq 0.9$ and $J_c \geq 1.4$; within this range, the value of compensation temperature increases as we increase the values of $J_s$ and $J_c$. For the critical temperature, we notice a significant increase in its value as we increase either $J_s$ or $J_c$ in the range of $J_s \leq 0.9$ and $J_c \geq 1.4$ where the system does not possess a compensation temperature.

The effect of the exchange and coupling parameter $J_{int}$ on the compensation and critical temperature is quite apparent in the results. We observe that the system only possesses a compensation temperature for $J_{int} \geq -0.3$; in addition, we observe that both critical and compensation temperatures increase as $J_{int}$ decreases.

The crystal field $D_s$ effect on the compensation and critical temperature of the system has been explored thoroughly in this work as well. As our results suggest, the compensation temperature is more affected by the variation of the crystal field $D_s$ than the critical temperature. We observe that the system exhibits a compensation temperature only for a specific range of $D_s$, which mainly depends on the values of the exchange and interaction parameters $J_s$ and $J_c$. For the range of $D_s$ in which the system exhibit compensation temperature, the value of the compensation temperature increases significantly with increasing $D_s$. The system critical temperature increases with increasing $D_s$ as well, however, the increase of the critical temperature occurs for specific values of $J_s$ and $J_c$, and its less significantly affected by the increase in $D_s$ than the compensation temperature.

The calculated values of the critical and compensation temperatures have been confirmed by the double peak phenomena observed in the susceptibilities and the specific heat curves. The first peak location coincides with the location of the compensation temperature and the second peak location coincides with the location of the critical temperature. We have noticed that the compensation temperature, in particular, is strongly linked to the Hamiltonian parameters.
The presence of the compensation temperature in our system makes it a potential candidate for thermomagnetic data storage and magneto-optical recording media devices [70, 71].

The effect of the Hamiltonian parameters on the system magnetic hysteresis loops has been studied thoroughly as well. The system temperature effect on these loops is obvious in the results as we notice a decrease in the loop's area as the temperature increases; also, we notice the system transition to a superparamagnetic state occurs at a high temperature of $T = 1.1$. The effect of the exchange and coupling parameters $J_s$ and $J_c$ on the hysteresis loops is quite apparent in the results as well, in which we notice an increase in the loop's area as the values of these parameters increase. In addition, we notice the appearance of multiple steps hysteresis loops with increasing the $J_s$ value. The exchange and coupling parameter $J_{int}$ seem to significantly affect the hysteresis loops since we notice that the loop's area increases as the value of $J_{int}$ decrease. Also, we notice the appearance of triple steps loops at a very low value of $J_{int}$. The effect of the crystal field $D_s$ on the hysteresis loops is obvious in the results, as well as we observe an increase in the loop's area as the value $D_s$ increases. In addition, we notice the appearance of two steps hysteresis loops at large values of $D_s$. The multiple hysteresis loop behaviors observed for our system are of great interest in the applications of multi-state memory devices [72].

**Acknowledgment**

This work was supported by the Jordan University of Science and Technology under Grant (20200039), and Khalifa University under grant CIRA-2019-043.

**References**

[1] Skomski, R. (2003). Nanomagnetics. Journal of physics: Condensed matter, 15(20), R841.

[2] Sellmyer, D. J., Zheng, M., & Skomski, R. (2001). Magnetism of Fe, Co and Ni nanowires in self-assembled arrays. *Journal of Physics: Condensed Matter*, *13*(25), R433.

[3] Bauer, L. A., Birenbaum, N. S., & Meyer, G. J. (2004). Biological applications of high aspect ratio nanoparticles. *Journal of Materials Chemistry*, *14*(4), 517-526.

[4] Tartaj, P. (2006). Nanomagnets-From fundamental physics to biomedicine. *Current Nanoscience*, *2*(1), 43-53.

[5] Bruck, E. (Ed.). (2017). *Handbook of Magnetic Materials*. Elsevier.

[6] Kodama, R. H. (1999). Magnetic nanoparticles. *Journal of magnetism and magnetic materials*, *200*(1-3), 359-372.

[7] López-Ortega, A., Tobia, D., Winkler, E., Golosovsky, I. V., Salazar-Alvarez, G., Estradé, S., ... & Nogués, J. (2010). Size-dependent passivation shell and magnetic properties in antiferromagnetic/ferrimagnetic core/shell MnO nanoparticles. *Journal of the American Chemical Society*, *132*(27), 9398-9407.

[8] López-Ortega, A., Estrader, M., Salazar-Alvarez, G., Roca, A. G., & Nogués, J. (2015). Applications of exchange coupled bi-magnetic hard/soft and soft/hard magnetic core/shell nanoparticles. *Physics Reports*, *553*, 1-32.

[9] B. Gleich and J. Weizenecker, "Tomographic imaging using the nonlinear response of magnetic particles," *Nature,* vol. 435, pp. 1214-1217, 2005.

[10] T. Hyeon, "Chemical synthesis of magnetic nanoparticles," *Chemical Communications,* pp. 927-934, 2003.

[11] S. Nie and S. R. Emory, "Probing single molecules and single nanoparticles by surface-enhanced Raman scattering," *science,* vol. 275, pp. 1102-1106, 1997.

[12] Pankhurst, Q. A., Connolly, J., Jones, S. K., & Dobson, J. (2003). Applications of magnetic nanoparticles in biomedicine. *Journal of physics D: Applied physics*, *36*(13), R167.

[13] Dai, Q., Berman, D., Virwani, K., Frommer, J., Jubert, P. O., Lam, M., ... & Nelson, A. (2010). Self-assembled ferrimagnet− polymer composites for magnetic recording media. *Nano letters*, *10*(8), 3216-3221.

[14] Zeng, H., Li, J., Liu, J. P., Wang, Z. L., & Sun, S. (2002). Exchange-coupled nanocomposite magnets by nanoparticle self-assembly. *Nature*, *420*(6914), 395-398.

[15] Koch, R. H., Deak, J. G., Abraham, D. W., Trouilloud, P. L., Altman, R. A., Lu, Y., ... & Parkin, S. S. P. (1998). Magnetization reversal in micron-sized magnetic thin films. *Physical review letters*, *81*(20), 4512.

[16] Emerich, D. F., & Thanos, C. G. (2003). Nanotechnology and medicine. *Expert opinion on biological therapy*, *3*(4), 655-663.

[17] Kurlyandskaya, G. V., Sanchez, M. L., Hernando, B., Prida, V. M., Gorria, P., & Tejedor, M. (2003). Giant-magnetoimpedance-based sensitive element as a model for biosensors. *Applied physics letters*, *82*(18), 3053-3055.

[18] Soong, R. K., Bachand, G. D., Neves, H. P., Olkhovets, A. G., Craighead, H. G., & Montemagno, C. D. (2000). Powering an inorganic nanodevice with a biomolecular motor. *Science*, *290*(5496), 1555-1558.

[19] Nielsch, K., Bachmann, J., Daub, M., Jing, J., Knez, M., Gösele, U., ... & Altbir, D. (2007). Ferromagnetic nanostructures by atomic layer deposition: from thin films towards core-shell nanotubes. *ECS Transactions*, *11*(7), 139.

[20] Armelles, G., Cebollada, A., García-Martín, A., Montero-Moreno, J. M., Waleczek, M., & Nielsch, K. (2012). Magneto-optical Properties of Core–Shell Magneto-plasmonic Au–Co x Fe3–x O4 Nanowires. *Langmuir*, *28*(24), 9127-9130.

[21] Coll, M., Montero Moreno, J. M., Gazquez, J., Nielsch, K., Obradors, X., & Puig, T. (2014). Low temperature stabilization of nanoscale epitaxial spinel ferrite thin films by atomic layer deposition. *Advanced Functional Materials*, *24*(34), 5368-5374.

[22] Kocakaplan, Y., & Keskin, M. (2014). Hysteresis and compensation behaviors of spin-3/2 cylindrical Ising nanotube system. *Journal of Applied Physics*, *116*(9), 093904.

[23] Keskin, M., Şarlı, N., & Deviren, B. (2011). Hysteresis behaviors in a cylindrical Ising nanowire. *Solid state communications*, *151*(14-15), 1025-1030.

[24] Zaim, A., Kerouad, M., Boughrara, M., Ainane, A., & de Miguel, J. J. (2012). Theoretical investigations of hysteresis Loops of ferroelectric or ferrielectric nanotubes with core/shell morphology. *Journal of superconductivity and novel magnetism*, *25*(7), 2407-2414.

[25] Masrour, R., Bahmad, L., Hamedoun, M., Benyoussef, A., & Hlil, E. K. (2013). The magnetic properties of a decorated Ising nanotube examined by the use of the Monte Carlo simulations. *Solid state communications*, *162*, 53-56.

[26] Konstantinova, E. (2008). Theoretical simulations of magnetic nanotubes using Monte Carlo method. *Journal of Magnetism and Magnetic Materials*, *320*(21), 2721-2729.

[27] Zaim, A., & Kerouad, M. (2010). Monte Carlo simulation of the compensation and critical behaviors of a ferrimagnetic core/shell nanoparticle Ising model. *Physica A: Statistical Mechanics and its Applications*, *389*(17), 3435-3442.

[28] Masrour, R., Bahmad, L., Hamedoun, M., Benyoussef, A., & Hlil, E. K. (2013). The magnetic properties of a decorated Ising nanotube examined by the use of the Monte Carlo simulations. *Solid state communications*, *162*, 53-56.

[29] S.-i. Ohkoshi and K. Hashimoto, "New magnetic functionalities presented by Prussian blue analogues," *Interface-Electrochemical Society,* vol. 11, pp. 34-38, 2002.

[30] B. Deviren and M. Keskin, "Dynamic phase transitions and compensation temperatures in a mixed spin-3/2 and spin-5/2 Ising system," *Journal of Statistical Physics,* vol. 140, pp. 934-947, 2010.

[31] G. Buendía and J. Villarroel, "Compensation temperatures of mixed ferro-ferrimagnetic ternary alloys," *Journal of Magnetism and Magnetic Materials,* vol. 310, pp. e495-e497, 2007.

[32] G. Chern, L. Horng, W. Shieh, and T. Wu, "Antiparallel state, compensation point, and magnetic phase diagram of Fe 3 O 4/M n 3 O 4 superlattices," *Physical Review B,* vol. 63, p. 094421, 2001.

[33] B. Boechat, R. Filgueiras, L. Marins, C. Cordeiro, and N. Branco, "Ferrimagnetism in two-dimensional mixed-spin Ising model," *Modern Physics Letters B,* vol. 14, pp. 749-758, 2000.

[34] A. Feraoun, A. Zaim, and M. Kerouad, "Monte Carlo study of a mixed spin (1, 3/2) ferrimagnetic nanowire with core/shell morphology," *Physica B: Condensed Matter,* vol. 445, pp. 74-80, 2014.

[35] A. Jabar and R. Masrour, "Spin compensation temperatures in the Monte Carlo Study of a Mixed spin-3/2 and spin-1/2 Ising ferrimagnetic system," *Journal of Superconductivity and Novel Magnetism,* vol. 30, pp. 2829-2834, 2017.

[36] A. Jabar, R. Masrour, A. Benyoussef, and M. Hamedoun, "Magnetic properties of the mixed spin-1 and spin-3/2 Ising system on a bilayer square lattice: A Monte Carlo study," *Chemical Physics Letters,* vol. 670, pp. 16-21, 2017.

[37] M. Karimou, R. Yessoufou, and F. Hontinfinde, "Critical behaviors and phase diagrams of the mixed spin-1 and spin-7/2 Blume-Capel (BC) Ising model on the Bethe lattice," *International Journal of Modern Physics B,* vol. 29, p. 1550194, 2015.

[38] A. A. Obeidat, M. K. Hassan, and M. H. Badarneh, "Magnetic properties and critical and compensation temperatures in mixed spin-1/2–spin-1 ferrimagnetic-centered rectangular structure using Monte Carlo simulation," *IEEE Transactions on Magnetics,* vol. 55, pp. 1-5, 2019.

[39] L. Gálisová and J. Strečka, "Magnetic and magnetocaloric properties of the exactly solvable mixed-spin Ising model on a decorated triangular lattice in a magnetic field," *Physica E: Low-dimensional Systems and Nanostructures,* vol. 99, pp. 244-253, 2018.

[40] M. Gharaibeh, A. Obeidat, M.-K. Qaseer, and M. Badarneh, "Compensation and critical behavior of Ising mixed spin (1-1/2-1) three layers system of cubic structure," *Physica A: Statistical Mechanics and its Applications,* vol. 550, p. 124147, 2020.

[41] B. Deviren, M. Ertaş, and M. Keskin, "The effective-field theory studies of critical phenomena in a mixed spin-1 and spin-2 Ising model on honeycomb and square lattices," *Physica A: Statistical Mechanics and its Applications,* vol. 389, pp. 2036-2047, 2010.

[42] Z. Elmaddahi and M. El Hafidi, "Magnetic properties of a three-walled mixed-spin nanotube," *Journal of Magnetism and Magnetic Materials,* vol. 523, p. 167565, 2021.

[43] R. Masrour, L. Bahmad, M. Hamedoun, A. Benyoussef, and E. Hlil, "The magnetic properties of a decorated Ising nanotube examined by the use of the Monte Carlo simulations," *Solid state communications,* vol. 162, pp. 53-56, 2013.

[44] R. Mendes, F. S. Barreto, and J. Santos, "Magnetic properties of the mixed spin 1/2 and spin 1 hexagonal nanotube system: Monte Carlo simulation study," *Journal of Magnetism and Magnetic Materials,* vol. 471, pp. 365-369, 2019.

[45] Kaneyoshi, T. (2011). Magnetic properties of a cylindrical Ising nanowire (or nanotube). *physica status solidi (b)*, *248*(1), 250-258.

[46] Kaneyoshi, T. (2011). Compensation temperature in a cylindrical Ising nanowire (or nanotube). *Physica A: Statistical Mechanics and its Applications*, *390*(21-22), 3697-3703.

[47] Kaneyoshi, T. (2005). Phase diagrams of a nanoparticle described by the transverse Ising model. *physica status solidi (b)*, *242*(14), 2938-2948.

[48] Elmaddahi, Z., & El Hafidi, M. (2021). Magnetic properties of a three-walled mixed-spin nanotube. *Journal of Magnetism and Magnetic Materials*, *523*, 167565.

[49] Masrour, R., Bahmad, L., Hamedoun, M., Benyoussef, A., & Hlil, E. K. (2013). The magnetic properties of a decorated Ising nanotube examined by the use of the Monte Carlo simulations. *Solid state communications*, *162*, 53-56.

[50] Mendes, R. G. B., Barreto, F. S., & Santos, J. P. (2019). Magnetic properties of the mixed spin 1/2 and spin 1 hexagonal nanotube system: Monte Carlo simulation study. *Journal of Magnetism and Magnetic Materials*, *471*, 365-369.

[51] Wang, W., Liu, Y., Gao, Z. Y., Zhao, X. R., Yang, Y., & Yang, S. (2018). Compensation behaviors and magnetic properties in a cylindrical ferrimagnetic nanotube with core-shell structure: A Monte Carlo study. *Physica E: Low-dimensional Systems and Nanostructures*, *101*, 110-124.

[52] Yang, Y., Wang, W., Ma, H., Li, Q., Gao, Z. Y., & Huang, T. (2019). Magnetic and thermodynamic properties of a ferrimagnetic mixed-spin (1/2, 1, 3/2) Ising nanoisland: Monte Carlo study. *Physica E: Low-dimensional Systems and Nanostructures*, *108*, 358-371.

[53] Kühl, F. G., Kazdal, T. J., Lang, S., & Hampe, M. J. (2017). Adsorption of sulfur dioxide and mixtures with nitrogen at carbon nanotubes and graphene: molecular dynamics simulation and gravimetric adsorption experiments. *Adsorption*, *23*(2-3), 293-301.

[54] Fang, Y. H., Tang, X. T., Sun, X., Zhang, Y. F., Zhao, J. W., Yu, L. M., ... & Zhao, X. L. (2017). Preparation and enhanced microwave absorption properties of Ni-Co attached single-walled carbon nanotubes and $CoFe_2O_4$ nanocomposites. *Journal of Applied Physics*, *121*(22), 224301.

[55] Wang, Z., Zhang, H., Cao, H., Wang, L., Wan, Z., Hao, Y., & Wang, X. (2017). Facile preparation of ZnS/CdS core/shell nanotubes and their enhanced photocatalytic performance. *International Journal of Hydrogen Energy*, *42*(27), 17394-17402.

[56] Rani, K. K., Devasenathipathy, R., Wang, S. F., & Yang, C. (2017). Simple preparation of birnessite-type $MnO_2$ nanoflakes with multi-walled carbon nanotubes for the sensitive detection of hydrogen peroxide. *Ionics*, *23*(11), 3219-3226.

[57] N. Espriella Velez, C. Ortega Lopez, and F. Torres Hoyos, "Critical and compensation temperatures for the mixed spin-3/2 and spin-5/2 Ising model," *Revista mexicana de física,* vol. 59, pp. 95-101, 2013.

[58] V. S. Leite, M. Godoy, and W. Figueiredo, "Finite-size effects and compensation temperature of a ferrimagnetic small particle," *Physical Review B,* vol. 71, p. 094427, 2005.

[59] H. P. D. Shieh and M. H. Kryder, "Magneto‐optic recording materials with direct overwrite capability," *Applied physics letters,* vol. 49, pp. 473-474, 1986.

[60] N. Metropolis, A. W. Rosenbluth, M. N. Rosenbluth, A. H. Teller, and E. Teller, "Equation of state calculations by fast computing machines," *The journal of chemical physics,* vol. 21, pp. 1087-1092, 1953.

[61] M. Newman and G. Barkema, *Monte carlo methods in statistical physics chapter 1-4* vol. 24: Oxford University Press: New York, USA, 1999.

[62] Wang, W., Chen, D. D., Lv, D., Liu, J. P., Li, Q., & Peng, Z. (2017). Monte Carlo study of magnetic and thermodynamic properties of a ferrimagnetic Ising nanoparticle with hexagonal core-shell structure. *Journal of Physics and Chemistry of Solids*, *108*, 39-51.

[63] Lv, D., Wang, F., Liu, R. J., Xue, Q., & Li, S. X. (2017). Monte Carlo study of magnetic and thermodynamic properties of a ferrimagnetic mixed-spin (1, 3/2) Ising nanowire with hexagonal core-shell structure. *Journal of Alloys and Compounds*, *701*, 935-949.

[64] Lv, D., Jiang, W., Ma, Y., Gao, Z., & Wang, F. (2019). Magnetic and thermodynamic properties of a cylindrical ferrimagnetic Ising nanowire with core/shell structure. *Physica E: Low-dimensional Systems and Nanostructures*, *106*, 101-113.

[65] Wang, W., Peng, Z., Lin, S. S., Li, Q., Lv, D., & Yang, S. (2018). Monte Carlo simulation of magnetic properties of a ferrimagnetic nanoisland with hexagonal prismatic core-shell structure. *Superlattices and Microstructures*, *113*, 178-193.

[66] Jiang, W., Li, X. X., Guo, A. B., Guan, H. Y., Wang, Z., & Wang, K. (2014). Magnetic properties and thermodynamics in a metallic nanotube. *Journal of magnetism and magnetic materials*, *355*, 309-318.

[67] Maaouni, N., Fadil, Z., Mhirech, A., Kabouchi, B., Bahmad, L., & Benomar, W. O. (2020). Compensation behavior of an anti-ferrimagnetic core-shell nanotube like-structure: Monte Carlo Study. *Solid State Communications*, *321*, 114047.

[68] Hachem, N., Badrour, I. A., El Antari, A., Lafhal, A., Madani, M., & El Bouziani, M. (2021). Phase diagrams of a mixed-spin hexagonal Ising nanotube with core-shell structure. *Chinese Journal of Physics*, *71*, 12-21.

[69] Liu, Y., Wang, W., Lv, D., Zhao, X. R., Huang, T., & Wang, Z. Y. (2018). Hysteresis behaviors in a ferrimagnetic Ising nanotube with hexagonal core-shell structure. *Physica B: Condensed Matter*, *541*, 79-88.

[70] R. Allen and G. Connell, "Magneto‐optic properties of amorphous terbium‐iron," *Journal of Applied Physics,* vol. 53, pp. 2353-2355, 1982.

[71] C. Savage, F. Marquis, M. Watson, and P. Meystre, "Direct overwrite in magneto‐optical recording," *Applied physics letters,* vol. 52, pp. 1277-1278, 1988.

[72] S. Bouhou, I. Essaoudi, A. Ainane, M. Saber, F. Dujardin, and J. de Miguel, "Hysteresis loops and susceptibility of a transverse Ising nanowire," *Journal of Magnetism and Magnetic Materials,* vol. 324, pp. 2434-2441, 2012.